\documentclass[aps,prb,superscriptaddress,showpacs,reprint]{revtex4-1}

\usepackage[utf8]{inputenc}
\usepackage{amsmath,amsfonts,amssymb}
\usepackage{pifont}
\usepackage{rotating}
\usepackage{enumerate}
\usepackage{graphicx}    
\usepackage{epstopdf}
\usepackage{color}
\usepackage{soul} 
\usepackage{subfigure}

\usepackage{multirow,bigstrut}
\newsavebox{\measurebox}

\usepackage{hyperref}
\definecolor{dark-red}{rgb}{0.9,0.15,0.15}
\definecolor{dark-blue}{rgb}{0.15,0.15,0.4}
\definecolor{medium-blue}{rgb}{0,0,0.5}
\hypersetup{
    colorlinks, linkcolor={black},
    citecolor={dark-blue}, urlcolor={medium-blue}
}

\begin{document}

\title{Co-existence of spin semi-metallic and Weyl semi-metallic behavior in FeRhCrGe}

\author{Y. Venkateswara}
\affiliation{Magnetic Materials Laboratory, Department of Physics, Indian Institute of Technology Bombay, Mumbai 400076, India}

\author{S. Shanmukharao Samatham}
\affiliation{Magnetic Materials Laboratory, Department of Physics, Indian Institute of Technology Bombay, Mumbai 400076, India}

\affiliation{Department of Physics, Maharaj Vijayaram Gajapathi Raj College of Engineering, Vijayaram Nagar Campus, Chintalavalasa, Vizianagaram 535005, Andhra Pradesh, India}

\author{P. D. Babu}
\affiliation{UGC-DAE Consortium for Scientific Research, Mumbai Centre, BARC Campus, Mumbai 400085, India}

\author{K. G. Suresh}
\email{suresh@phy.iitb.ac.in}
\affiliation{Magnetic Materials Laboratory, Department of Physics, Indian Institute of Technology Bombay, Mumbai 400076, India}

\author{Aftab Alam}
\email{aftab@phy.iitb.ac.in}
\affiliation{Magnetic Materials Laboratory, Department of Physics, Indian Institute of Technology Bombay, Mumbai 400076, India}
\affiliation{Department of Physics, Indian Institute of Technology Bombay, Mumbai 400076, India}


\begin{abstract}

In this letter, we report the discovery of a new class of spintronic materials, namely spin semi-metals (SSM), employing both theoretical and experimental tools. The band structure of this class of materials is such that one of the spin bands resembles that of a semi-metal, while the other is similar to that of an insulator/semiconductor. This report is the experimental verification of the first SSM, FeRhCrGe, a quaternary Heusler alloy with a magnetic moment 3 $\mu_B$ and a Curie temperature of 550 K. The measurement below 300 K shows nearly temperature independent conductivity and a relatively moderate Hall effect. SSM behavior for FeRhCrGe is also confirmed by rigorous first principles calculations. Band structure calculations also reveal that the spin up (semi metallic) band has combined features of type II Weyl and nodal line semimetal. As such, this study opens up the possibility of a new class of material with combined spintronic and topological properties, which is important both from fundamental and applied point of view.

\end{abstract}


\date{\today}
\pacs{85.75.-d, 75.47.Np, 75.76.+j, 76.80.+y}  
\maketitle

{\bf{Introduction:}}
Recent developments in magnetic Heusler alloys, such as the discovery of  half metallic ferromagnets (HMF) have fuelled the area of spintronics research  considerably.\cite{deGroot-NiMnSb-HMF-prl-theory,Katsnelson-half-metals-RevModPhy,Felser-spintronics-ACIE-review,felser2013spintronics-book,ching2013half-book} One of the spin bands is metallic while the other is either semiconducting or insulating in this class of materials. Later, spin gapless semiconductors (SGS)\cite{XLWang-SGS-prl-theory} were discovered which gained prominence over the half metals due to their unique properties. Magnetic semiconductors\cite{Xingxing-bms-nanoscale} with unequal band gaps constitute another class of spintronic materials that produce high spin polarized charge carriers at elevated temperatures. Fully compensated ferrimagnet\cite{CrVTiAl-YSS-prb} is another unique class. In this letter, we discovered a new interesting class of materials, which we term as spin semi-metals (SSM). One of the main motives of this paper is to highlight the discovery of SSM and its importance and applications in the field of spintronics. We demonstrate the existence of necessary features of SSM in a realistic material,  FeRhCrGe. Furthermore, the spin up band of this material is also found to acquire a combined feature of type II Weyl\cite{Xu-science-Weyl,Huang-Weyl-PRX} and nodal line\cite{Fang-nodal-line-PRB,Bian-nodalline-nature} semi-metal.

One of the necessary conditions for a spintronic material is to possess integer magnetic moment ($M$), for which the density of states (DoS) should be zero for at least one spin bands.\cite{Venkatesan-HandMMM-halfmetals} This leads to six possible classes of materials namely (i) conventional semiconductors, (ii) simple gapless semiconductors, (iii) magnetic semiconductors, (iv) half metals, (v) SGS and (vi) SSM. Schematic band structures of all these are shown in Fig S1 of supplementary material (SM).\cite{supp} The first two classes  do not produce any spin polarized carriers in their electrical conduction due to symmetric spin up and down DoS. Spin semimetals are expected to be potential materials as they produce both spin polarized electrons and holes, similar to SGS materials. The identification of many HMF and SGS materials from the Heusler family has resulted in a renewed interest in the field of spintronics due to their interesting properties such as high Curie temperature (T$_C$), structural stability and compatibility to grow thin films.\cite{felser2015heusler-book,Bainsla-AppPhysRev} In this context,  we would like to highlight an interesting observation that, only a certain discrete set of lattice parameter of Heusler alloys gives rise to the above mentioned six classes of materials.  This critical separation is found to be nearly 0.123 \AA. Taking the lattice parameter of Co$_2$MnSi\cite{Raphael-Co2MnSi-prb} ($\sim$5.65 \AA) as one of the those allowed values, the lattice parameters corresponding to other Heusler alloys are tabulated in Table S1 of SM.\cite{supp}  

\begin{figure}[t!]
\includegraphics[width=\linewidth]{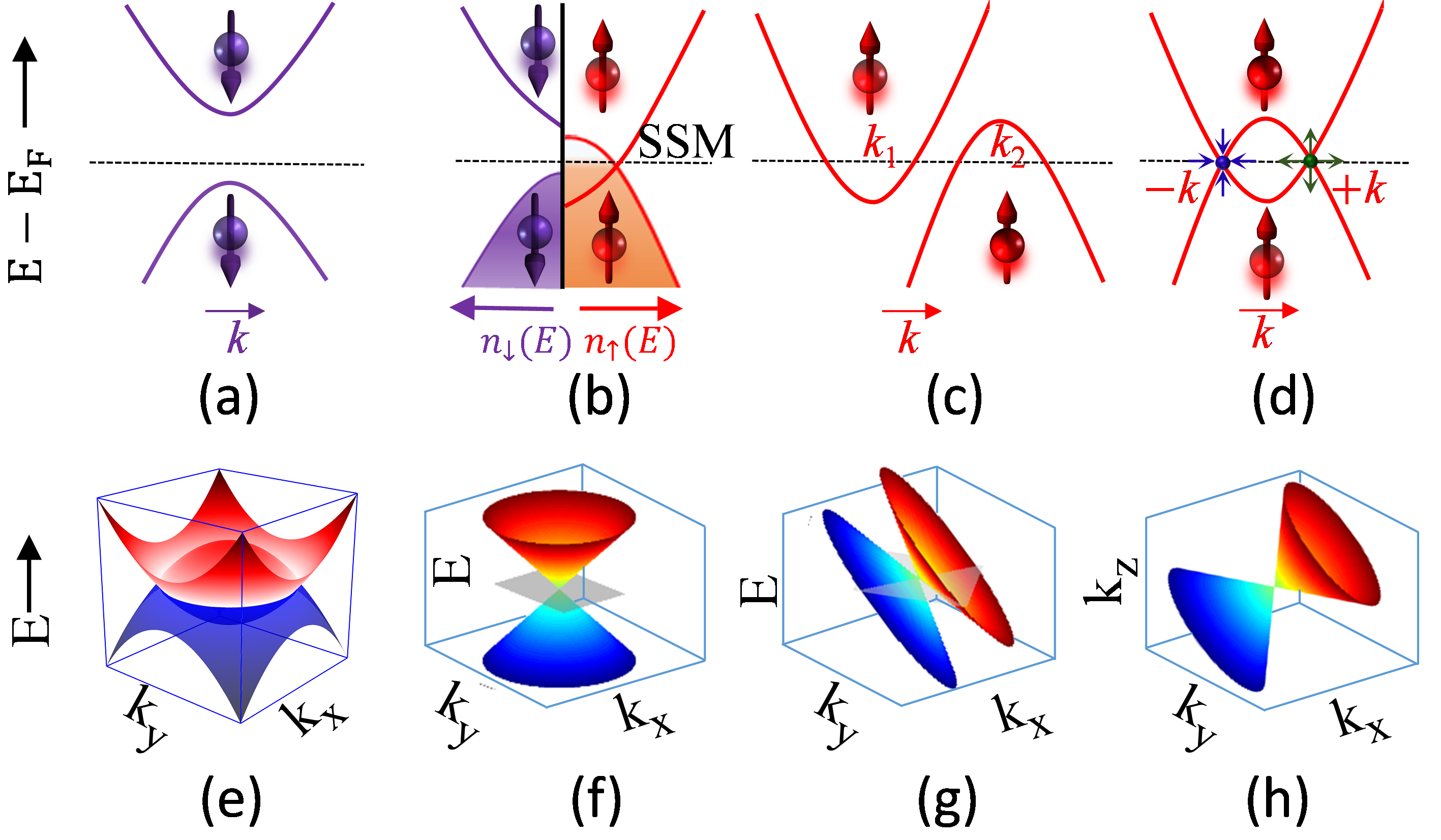} 
\caption{Schematics of spin resolved band structure of a SSM with possibilities of different types of topological nodes in spin up band. (a) Spin down band with a gap, (b) Spin resolved DoS. Spin up band with (c) indirect overlap of valence and conduction bands (d) direct overlap of valence and conduction bands resulting in topological Weyl points in 1D band. Schematic 2D projected spin up band structure of (e) nodal line (f) type I Weyl and (g) type II Weyl semimetals. (h) Fermi surface, as a Dirac cone, around type II Weyl point.}
\label{fig1}
\end{figure}

Spin semimetals are a unique class of spintronic materials, in which one of the spin bands is semi-metallic while the other is either insulating or semiconducting. The spin semimetallic band consists of partial overlap of two or more different types of valence and conduction bands. It causes a large DoS at the Fermi level (E$_\text{F}$) in the spin semimetallic band. However, they show negative temperature coefficient in their electrical resistivity, which is due to the presence of electron and hole pockets at E$_{\text{F}}$. Figure \ref{fig1}(b) shows a schematic spin resolved DoS for SSM, in which spin down DoS is gapped (bands shown in Fig.\ref{fig1}(a)), while the spin up DoS is due to the partial overlap of valence and conduction bands with the possibilities of indirect (Fig.\ref{fig1}(c)) or direct (Fig.\ref{fig1}(d)) overlapping. Direct overlap often causes  doubly degenerate  band crossing at ${\it{-k}}$ and ${\it{+k}}$  points, which are analogous to Weyl points in 1D. Weyl nodes or points are the points in the 3D Brillouin zone (BZ) which possess cone-like energy spectrum.\cite{Wan-Y2Ir2O7-prb-theory} It is well known that the Weyl nodes are protected in magnetic systems due to the lifting of spin degeneracy at the nodes.\cite{Wan-Y2Ir2O7-prb-theory,Wang-Co2ZrSn-magnWeyl-prl-theory} These nodes act as sources and sinks of Berry flux. In fact, the 1D Weyl like nodes can evolve as either (i) nodal line or (ii) type I or type II Weyl points in the 3D band structure. 2D projected band structure for nodal line, type I and type II Weyl points are shown in Fig. \ref{fig1}(e)-(g). It has been reported that the Weyl nodes can occur at points other than high symmetry $k$-path unlike nodal line semimetals.\cite{Wan-Y2Ir2O7-prb-theory,Shuo-Ying-APX-review-nodaline}  Type I Weyl points cause point-like Fermi surface (FS) and hence difficult to identify. However, 2D band structure of tilted Dirac cone causes the crossed Fermi lines (see Fig.\ref{fig1}(g)) centred at Weyl node. \cite{Soluyanov-IIweyl-natureLett-theory} These crossed Fermi lines evolve into Dirac cone centred at type II Weyl point in the 3D band structure (see Fig.\ref{fig1}(h)). Hence, the presence of type II Weyl points can be easily identified by locating the position of Dirac cone FS in the 3d BZ. To be a Weyl point, each cone must belong to different bands.\cite{Zubkov-IIweyl-simulated-theory}

In general, the nature of overlapping bands near E$_{\text{F}}$ can be different e.g. overlap of either (i) flat valence and sharp conduction bands or (ii) sharp valence and flat conduction bands, or (iii) flat valence and flat conduction bands with E$(\mathbf{k})$ spectrum different from parabolic nature. Different possibilities of E$(\mathbf{k})$ spectrum near E$_{\text{F}}$ were discussed by X. L. Wang\cite{XLWang-SGS-prl-theory} for SGS. The same scenario can be expected in SSM as well. However, unlike SGS materials in which spin gapless nature is extremely sensitive to the disorder, pressure and impurities, SSM are expected to be less sensitive yet giving a high spin polarization due to the presence of large DoS at $E_F$.  With the semimetallic nature of one spin band in SSM, this also  leads to the possibility of hosting Weyl fermions in the semimetallic band, in addition to the presence of standard spin polarized electrons and holes. As such, SSM materials having signatures of Weyl nodes lead to a completely new era for spintronic materials.

In this letter, we report the first spin semimetallic material, namely quaternary Heusler alloy FeRhCrGe, based on the confirmation from both experimental and theoretical investigations. We also confirm that this alloy shows type II Weyl nodes in the semimetallic band, resulting in the realization of Weyl fermions.\cite{Shi-Ti2MnAl-prbRapid-theory} Finally, with the help of molecular energy level diagram, we propose that, 27 valence electron quaternary alloys can, in general, be a  promising playground to find potential SSM materials.

FeRhCrGe belongs to a quaternary Heusler alloy (XX$^{'}$YZ) with space group F$\bar{4}3$m. The structure can be seen as 4 interpenetrating fcc sublattices with Wyckoff positions 4a, 4b, 4c and 4d.  The possible non-degenerate configurations for a general XX$'$YZ quaternary alloy (fixing Z-atom at 4a-site) are
(I) X at 4c, X$'$ at 4d and Y at 4b site,
(II) X at 4b, X$'$ at 4d and Y at 4c site,
 (III) X at 4c, X$'$ at 4b and Y at 4d site.
 Configuration I turns out to be energetically the most stable one.

\section{Experimental results}

\begin{figure}[b!]
\centering
\includegraphics[width=0.8\linewidth]{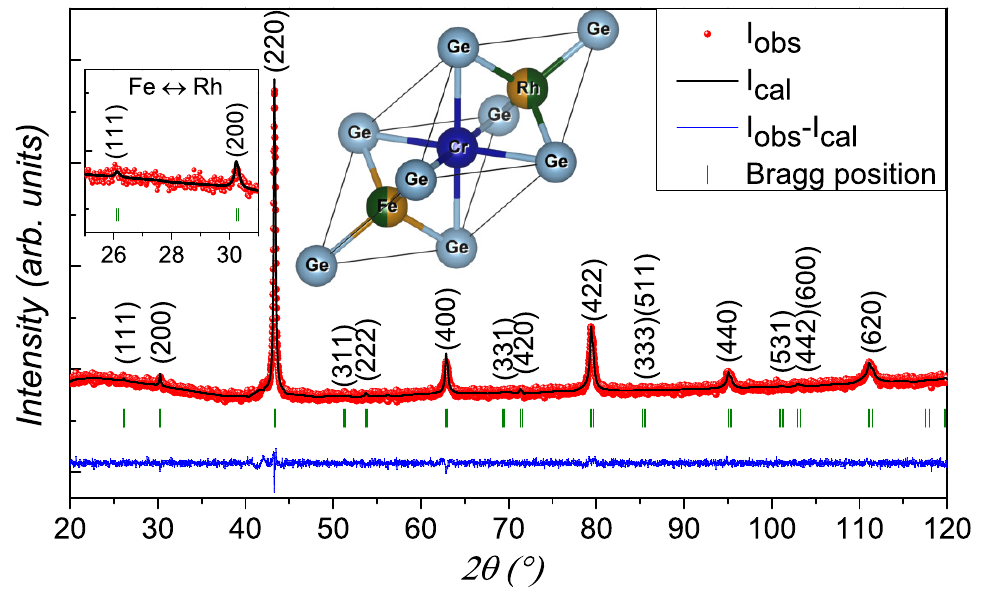}
\caption{Room temperature XRD pattern of FeRhCrGe along with the Rietveld refinement corresponding to the crystal structure shown in the inset. Inset (left corner) shows the zoomed-in view of XRD pattern near (111) and (200) peaks, confirming the L2$_1$ disorder in the alloy.}
\label{fig:xrd-ferhcrge}
\end{figure}

{\it \bf Crystal Structure : } Room temperature X-ray diffraction (XRD) pattern along with the Rietveld refinement for the configuration I with 50\% anti-site disorder between tetrahedral site atoms Fe and Rh is shown in Fig. \ref{fig:xrd-ferhcrge}. A more detailed analysis of XRD pattern is presented in SM.\cite{supp} Due to the presence of antisite disorder, we conclude that FeRhCrGe crystallizes in L2$_1$ structure (crystal structure shown in the inset of Fig. \ref{fig:xrd-ferhcrge}).

{\it \bf Magnetization :} According to the Slater-Pauling (SP) rule,\citep{SimpleRulesHeusler} the saturation magnetization ($M_S$ in units of $\mu_B$) in the present case is given by $M_S = N_v - 24$, where $N_v$ is the total number of valence electrons of the Heusler alloy. It predicts $M_S=$ 3 $\mu_B$ for FeRhCrGe.

Magnetization data (M vs. H and M vs. T) indicating ferromagnetic behaviour of FeRhCrGe is shown in Fig. S3 of SM.\cite{supp} $M_S$ at 3 K is nearly 2.9 $\mu_B/f.u.$, which agrees fairly well with  SP rule. T$_\text{C}$ is found to be $\sim$ 550 K.

\begin{figure}[t]
\centering
\includegraphics[width=\linewidth]{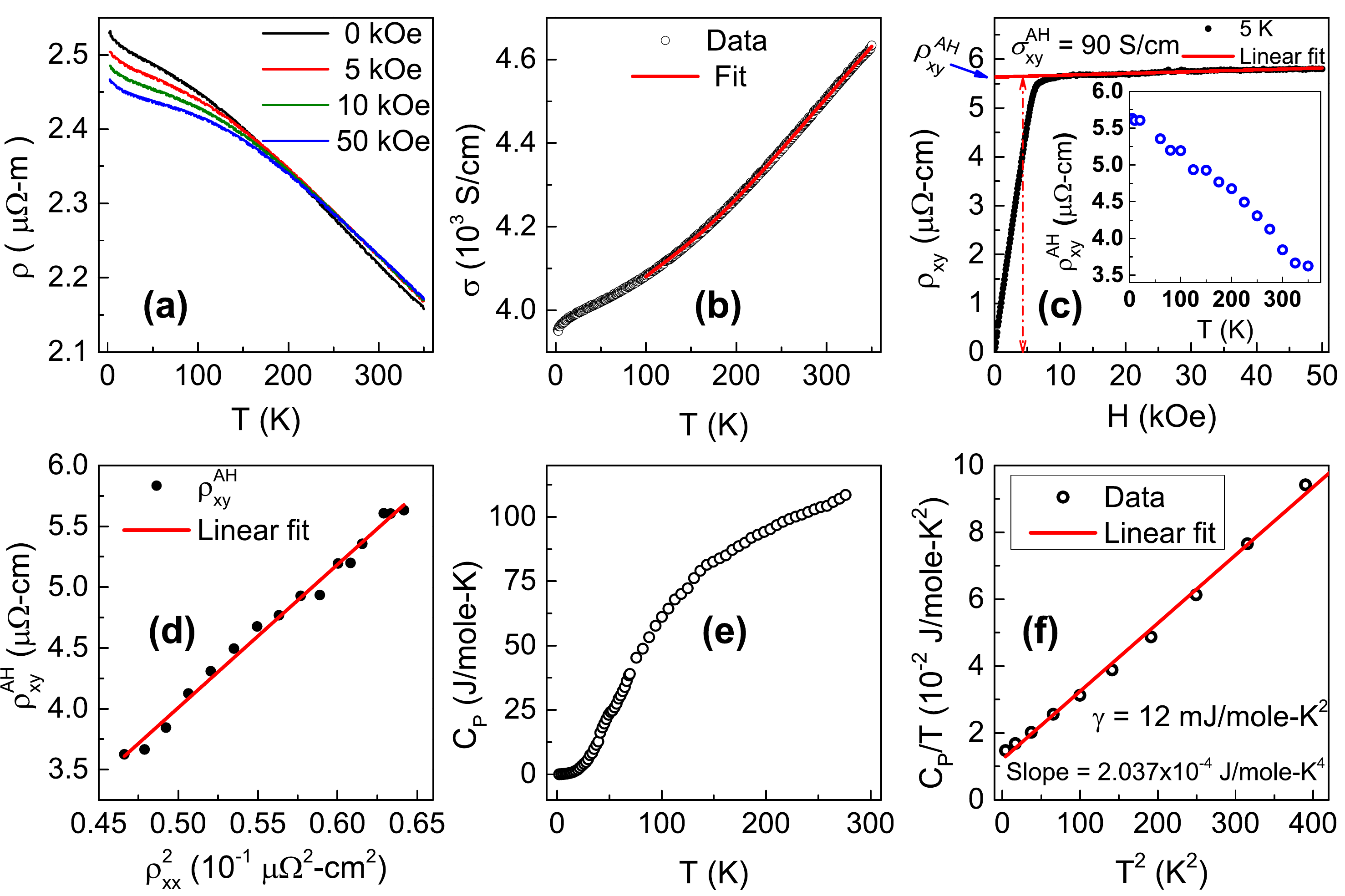} 
\caption{For FeRhCrGe (a) Resistivity ($\rho$) vs. T at different fields (H). (b) Two-carrier model fit to the conductivity data. (c) Hall resistivity ($\rho_{xy}$) vs. H at 5 K, and inset shows $\rho_{xy}$ vs. T. (d) $\rho_{xy}^{AH}$ vs. $\rho_{xx}^2$ along with the linear fit. (e) Heat capacity ($\mathrm{C_P}$) vs. T. (f) $\mathrm{C_P/T}$ vs. $\mathrm{T^2}$ and a linear fit to it in the range 2-20 K}
\label{fig:R-H-FeRhCrGe}
\end{figure}

{\it\bf Transport properties :} Figure \ref{fig:R-H-FeRhCrGe}(a) shows resistivity ($\rho$) vs. T at different H for FeRhCrGe. Negative temperature coefficient of $\rho$ indicates the possibility of either a semiconductor or a semimetal. However, the absence of exponential dependence precludes the gapped semiconducting nature. Therefore, it could either be a simple gapless semiconductor, SGS or a semimetal. To further understand the transport behavior, conductivity data is fitted with the modified two-carrier model\cite{kittel2007-conductivity,Mn3Al-pra} 
\begin{equation}
\sigma = e (n_e \mu_e + n_h \mu_h)
\label{eq:tbm}
\end{equation}
where, $n_i=n_{i0} e^{-\Delta E_i/k_\mathrm{B}T}$ $(i=e,h)$ are the carrier concentrations of electrons, holes with mobilities $\mu_i$. $\Delta E_i$ are the energy gaps. A fit to $\sigma$ in the zero field, high T-region is shown in Fig. \ref{fig:R-H-FeRhCrGe}(b). The energy gaps obtained are 53.4 and 0.3 meV. Using the Hall data (explained below), the lower band gap is assigned to the holes. 

Hall resistivity ($\rho_{xy}$) vs. H is shown in Fig. \ref{fig:R-H-FeRhCrGe}(c) which is typical of a ferromagnetic material. In such materials, $\rho_{xy}$ is contributed by both ordinary and anomalous Hall resistivities i.e., $\rho_{xy}= \rho_{xy}^{OH}+\rho_{xy}^{AH}= R_0 H+R_\mathrm{AH}M$, where  $R_{0}$ ($R_{AH}$) is the coefficient of ordinary (anomalous) Hall resistivity. The Hall conductivity ($\sigma_{xy}$) is estimated using the relation, $\sigma_{xy}=\rho_{xy}/({\rho_{xy}^2+\rho_{xx}^2})$

The estimated anomalous Hall conductivity $\sigma_{xy}^{AH}$ $\simeq$ 90 S/cm at 5 K. The carrier concentration was estimated from the ordinary Hall coefficient ($R_0$) which in turn is extracted from the linear fit of high field data (above 15 kOe). The positive value of the ordinary Hall coefficient hints to the fact that the holes are the majority charge carriers. The concentration of hole carriers is found to be $\sim 10^{22}/cc$. Anomalous Hall resistivity ($\rho_{xy}^\mathrm{AH}$) is estimated from the y-intercept of the extrapolated high field linear fit. The inset of Fig. \ref{fig:R-H-FeRhCrGe}(c) shows the T-dependence of $\rho_{xy}^\mathrm{AH}$, which follows a similar trend as that of zero-field resistivity. In order to obtain the intrinsic and extrinsic anomalous Hall conductivities, $\rho_{xy}^\mathrm{AH}$ is scaled to $\rho_{xx}^2$ (see Fig. \ref{fig:R-H-FeRhCrGe}(d)) using the empirical formula,\cite{Tian-prl-Scaling-AHE} 
$\rho_{xy}^\mathrm{AH} = a + b\rho_{xx}^2$.
The fitting gives $a = -1.8\ \mu\Omega.cm$ and $b \sim 118\ S/cm$, where $a$ and $b$ are related to extrinsic and intrinsic contributions, respectively. The negative value of $a$ indicates that the extrinsic contribution is opposite to the Karplus-Luttinger term\cite{Karplus-PR-theory} (intrinsic contribution). The intrinsic anomalous Hall conductivity is $\sigma^\mathrm{int}_{xy} = -b$.\cite{Tian-prl-Scaling-AHE}  We have also simulated the intrinsic anomalous Hall conductivity using Berry curvature calculations (see Sec. II). The measured $\sigma^\mathrm{int}_{xy}$ is nearly one fourth of the theoretically estimated value. Such a low value could be due to the presence of both types of charge carriers and degenerate flat bands at E$_{\text{F}}$, in which some of the heavy carriers get trapped and easily recombine with the opposite carriers before they reach the edges of the sample.

{\it\bf Specific heat :}
Figure \ref{fig:R-H-FeRhCrGe}(e) shows the specific heat (C$_{\text{P}}$) vs. T. The electronic contribution to C$_{\text{P}}$ in simple gapless semiconductor or SGS is quite small compared to that of metals or semi-metals due to vanishing DoS at E$_{\text{F}}$.  Figure \ref{fig:R-H-FeRhCrGe}(f) shows the $\mathrm{C_P/T}$ vs. T$^2$ along with the linear fit. Using the free electron gas model, the DoS at E$_\text{F}$ can be estimated from\cite{PhysRevB-Ru2TaAl-semimetal} $n(\text{E}_\text{F})=3\gamma/(\pi^2 k_B^2)$, where $\gamma$ is the Sommerfeld constant. The value of $\gamma$ (12 mJ/mole-K$^2$), obtained from the fit, yields $n(\text{E}_\text{F})\sim 5.05$ states/eV-f.u.,  which is in good agreement with the simulated results (see Sec. II). Hence, it is quite evident that FeRhCrGe is neither a simple gapless semiconductor nor a SGS; instead it is very likely to be a spin semimetal.

\section{Theoretical Results}

Computational details are given in Sec. C of SM.\cite{supp}
Within ab-initio framework, all the three non degenerate configurations were fully relaxed with parallel spin moments on each magnetic ions and the results are shown in Table II of SM.\cite{supp} Configuration I turns out to be the ground state with ferromagnetic ordering. This configuration is consistent with the empirical rule based on electronegativities.\citep{SimpleRulesHeusler,EYS-prb} 
Theoretically optimized lattice parameter ($a_0=5.78$ \AA) agrees well with experimental value. The calculated net magnetization ($3\ \mu_B$/f.u.) is in good agreement with the value obtained from SP rule, as well as that of experiment (2.90 $\mu_B/f.u.$) at 3 K.

\begin{figure}[b!]
\centering
\includegraphics[width=\linewidth]{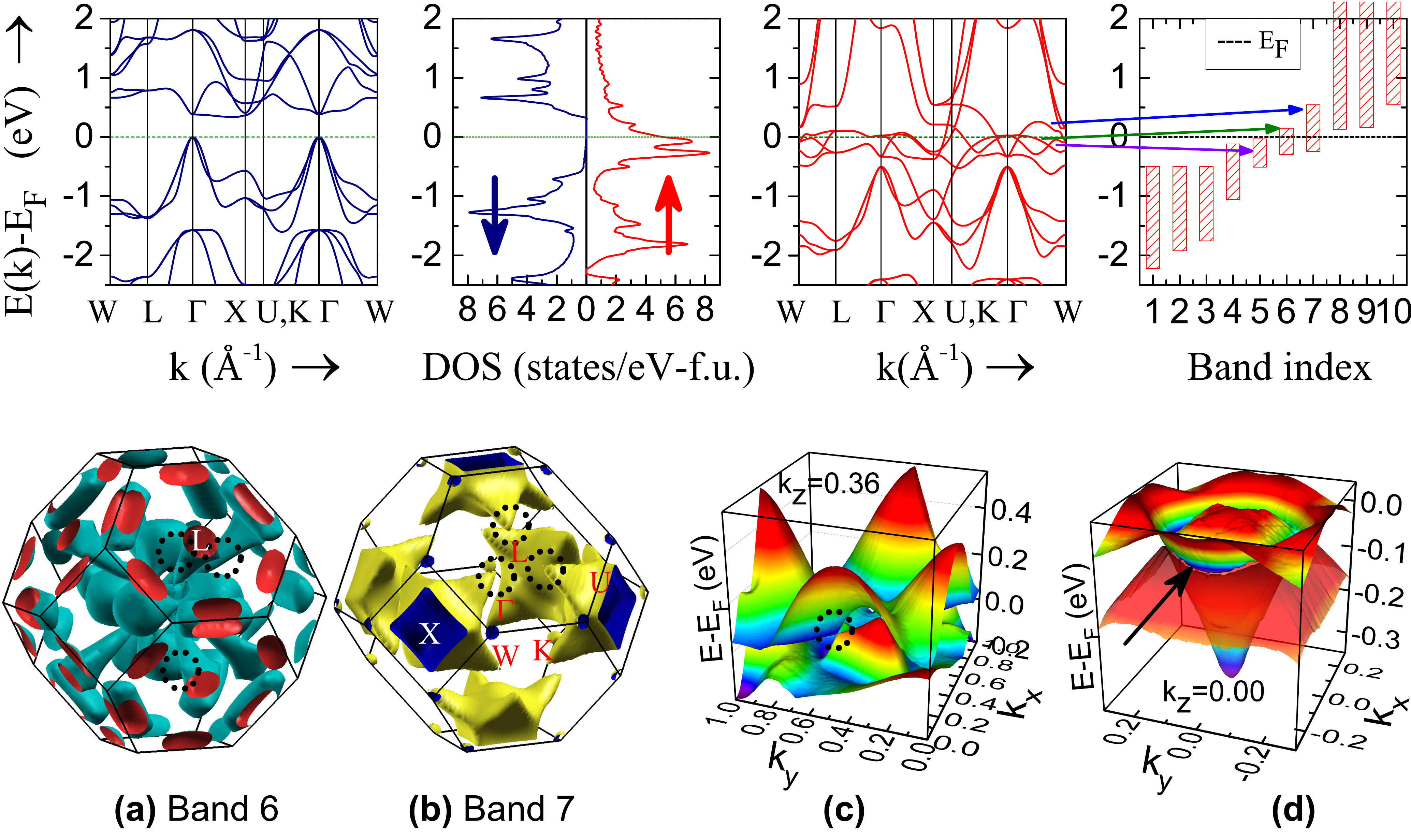} 
\caption{For FeRhCrGe, (top) spin polarized band structure and DoS. Extreme right panel shows the widths of various bands, illustrating the band flatness near E$_{\text{F}}$. 
(Bottom) Fermi surface arising due to (a) band 6 (b) band 7 in the spin up band. Dotted circles indicate the locations of Dirac cone FS. E$(\mathbf{k})$ spectrum at (c) $k_z=0.36\times (2\pi/a)$ plane having tilted Dirac cones due to spin up bands 6 \& 7 (shown by dotted circle) and (d) $k_z=0$ plane due to spin up bands 5 \& 6 causing the nodal line (arrow indicates location of nodal line).}
\label{fig:DOSBS-FeRhCrGe}
\end{figure}

Figure \ref{fig:DOSBS-FeRhCrGe} (top) shows the spin polarized band structure and DoS for the ground state configuration at the relaxed lattice parameter. The extreme right panel (top) shows the band width of various spin up bands contributing to the DOS near E$_{\text{F}}$. Notably, the large DoS at E$_{\text{F}}$ for spin up channel mainly arises from flat bands. Combination of bands 6 and 7 resembles that of a typical semi-metal. 
Figure \ref{fig:DOSBS-FeRhCrGe}(a,b) (bottom) shows the FS arising out of bands 6 and 7.
 As discussed earlier, type II Weyl nodes cause the Dirac cone FS at that point. The location of these points is indicated by dotted circles in Fig. \ref{fig:DOSBS-FeRhCrGe}(a,b). Figure \ref{fig:DOSBS-FeRhCrGe}(c) shows the E$(\mathbf{k})$ spectrum at $k_z=0.72(\pi/a)$ plane. The positions of tilted Dirac cones were found at (0.36,0.59,0.36)$\times 2\pi/a$ and (0.59,0.36,0.36)$\times 2\pi/a$. The other point exists at (0.64,0.64,0.41)$\times 2\pi/a$. There are a total of 24 type II Weyl points in the entire BZ (3  around each of 8 $L$-points). A similar scenario was also observed by Wan \textit{et. al.} in Y$_2$Ir$_2$O$_7$,\cite{Wan-Y2Ir2O7-prb-theory}  having type I Weyl points. Weyl points are the sources of 'Berry flux' in the momentum space but the net flux due to all Weyl points in the BZ vanishes (due to opposite chirality).\cite{Wan-Y2Ir2O7-prb-theory} To understand the pattern of Berry curvature around these points, non-collinear spin calculations with spin orbit coupling (SOC) were also performed.

\begin{figure}[t!]
\centering
\includegraphics[width=1.0\linewidth]{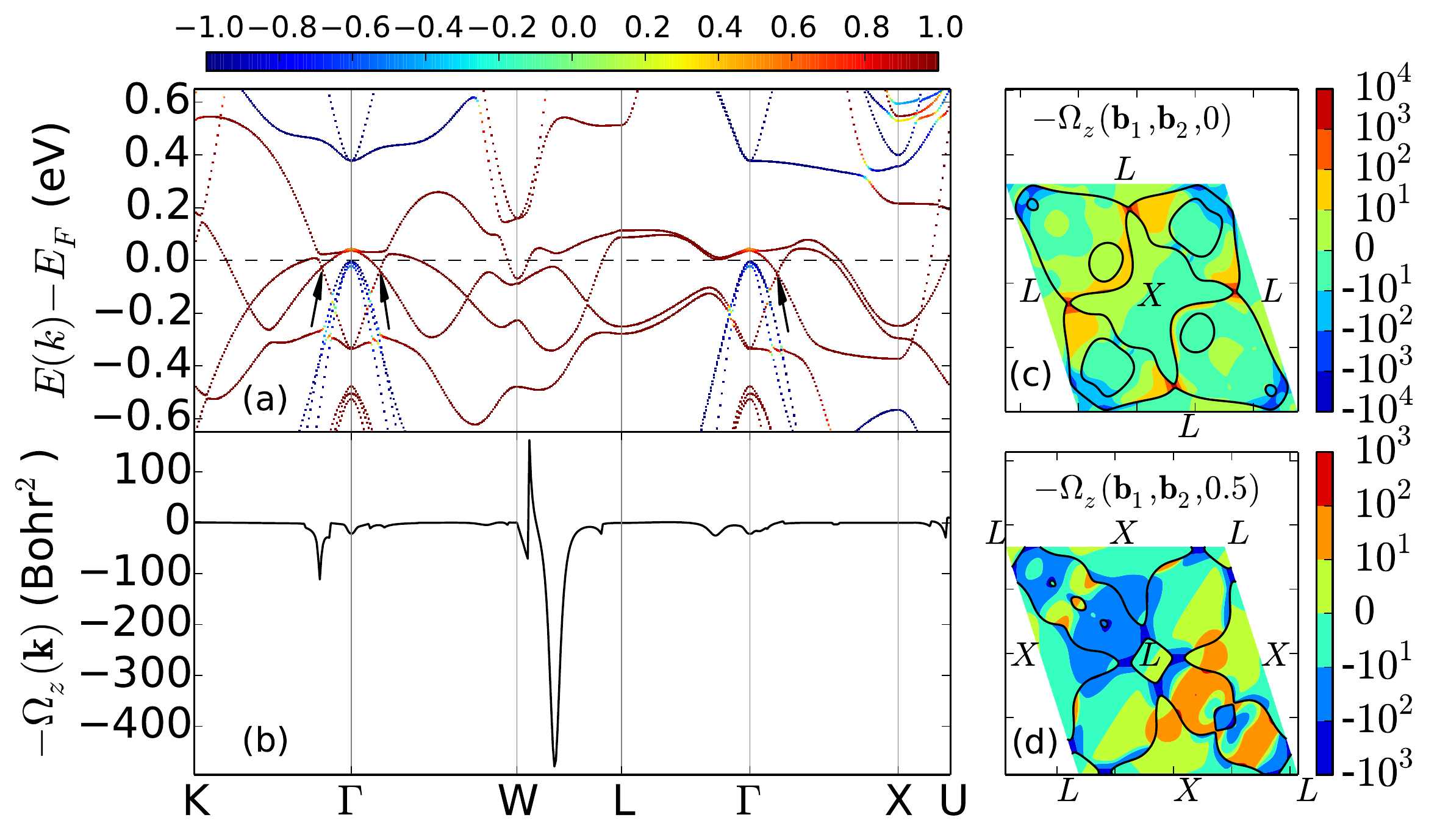}
\caption{For FeRhCrGe, (a) band structure including SOC with color profile showing the value of projected spin on the magnetization direction. (b) Berry curvature ($-\Omega_z(\bf k)$) along high symmetry lines. 2D projection of $-\Omega_z(\bf k)$ on (c) $\mathbf{b_3}=0$  and (d) $\mathbf{b_3}=0.5$ plane. Notice the change of magnitude/sign of Berry flux at Weyl like points near L-point as we move from $\mathbf{b_3}=0$ to $\mathbf{b_3}=0.5$ plane.}
\label{fig:BandBerry-curv}
\end{figure}

\begin{figure}[b!]
\centering
\includegraphics[width=\linewidth]{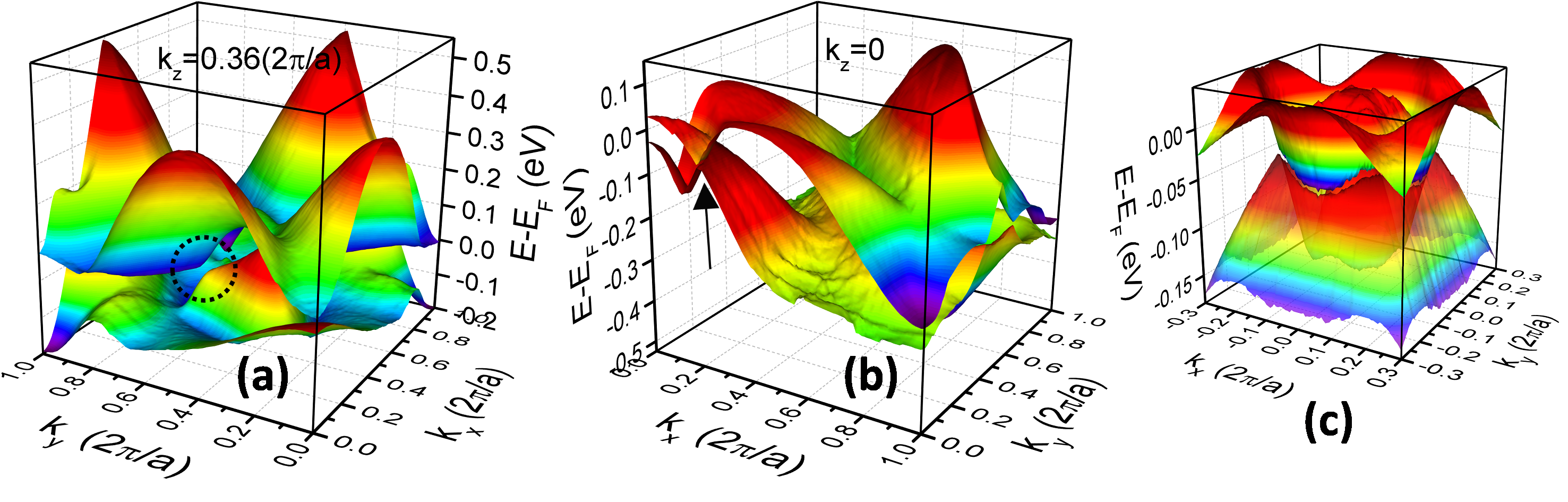} 
\caption{For FeRhCrGe, E$(\mathbf{k})$ spectrum including SOC effect showing (a) gapped topological Dirac node on $k_z=0.36$ and (b) nodal line semimetal on $k_z=0$ plane. Position of nodal line is shown by arrow. (c) Zoomed in view near nodal point. }
\label{fig:FeRhCrGe-nodal-SOC}
\end{figure}

With inclusion of  SOC, magnetic ground state of configuration I turns out to be collinear with total moment $3.0$ $\mu_B/f.u.$. The moments on individual magnetic ions are nearly same as those obtained from collinear spin calculations. Figure \ref{fig:BandBerry-curv}(a) shows the SOC band structure, with th colour profile indicating the value of projected spin on the magnetization direction. Apparently, spin character of the bands near E$_{\text{F}}$ is pure, which is crucial to retain high spin polarization and also spin semi-metallic nature. Figure \ref{fig:BandBerry-curv}(b) shows the z-component of the simulated Berry curvature $\Omega_z(\mathbf{k})$ along the high symmetry {\bf k}-points while Fig. \ref{fig:BandBerry-curv}(c) and (d) shows $\Omega_z(\mathbf{k})$ on $\mathbf{b_3}=0$ and $\mathbf{b_3}=0.5$ planes respectively. Notice the change of sign of Berry flux around L point, as we move from $\mathbf{b_3}=0$ to $\mathbf{b_3}=0.5$ planes. Magnitude of flux, however, is not same which needs to be equal and opposite to be a Weyl point.\cite{Wan-Y2Ir2O7-prb-theory} This indicates the deviation from a tilted Dirac cone, with the onset of a gap arising out of SOC effect. This is precisely shown in Fig. \ref{fig:FeRhCrGe-nodal-SOC}(a). The $e_g$ band (having spin up character) lying just below E$_{\text{F}}$ ($\sim 28$ meV), retains topological character. Notice from Fig. \ref{fig:BandBerry-curv}(a), that the degeneracy of these bands at certain points along high symmetry lines is not lifted, i.e., (i) one set of points along $\Gamma-K$, $\Gamma-W$ and $\Gamma-X$ indicating the possibility of nodal line semi-metal with energies lying between 28 to 50 meV below E$_{\text{F}}$ and (ii) few other nodal and Weyl points below $-0.1$ eV. Figure \ref{fig:FeRhCrGe-nodal-SOC}(b) shows the presence of nodal line in the spin up  band. A zoomed view around the nodal point is shown in Fig. \ref{fig:FeRhCrGe-nodal-SOC}(c). Very large Berry flux is seen at points (1,0.5,0.1)$\times 2\pi/a$, which are close to the high symmetry point $W$. E$(\mathbf{k})$ has Dirac cone like behavior at these points. The intrinsic anomalous Hall conductivity is obtained by integrating $\Omega_z(\mathbf{k})$ over the entire BZ,\cite{Yao-Berry-Fe-prl-theory} i.e., $\sigma_{xy}^{int}= - e^2/(8\pi^3\hbar)  \int_{BZ} d^3k\ \ \Omega_z(\mathbf{k})$, which turns out to be -495 $S/cm$. Obviously, this large value is due to the mismatch of the magnitude of Berry flux at the locations, as discussed earlier.

\begin{figure}[t!]
\centering
\includegraphics[width=0.7\linewidth]{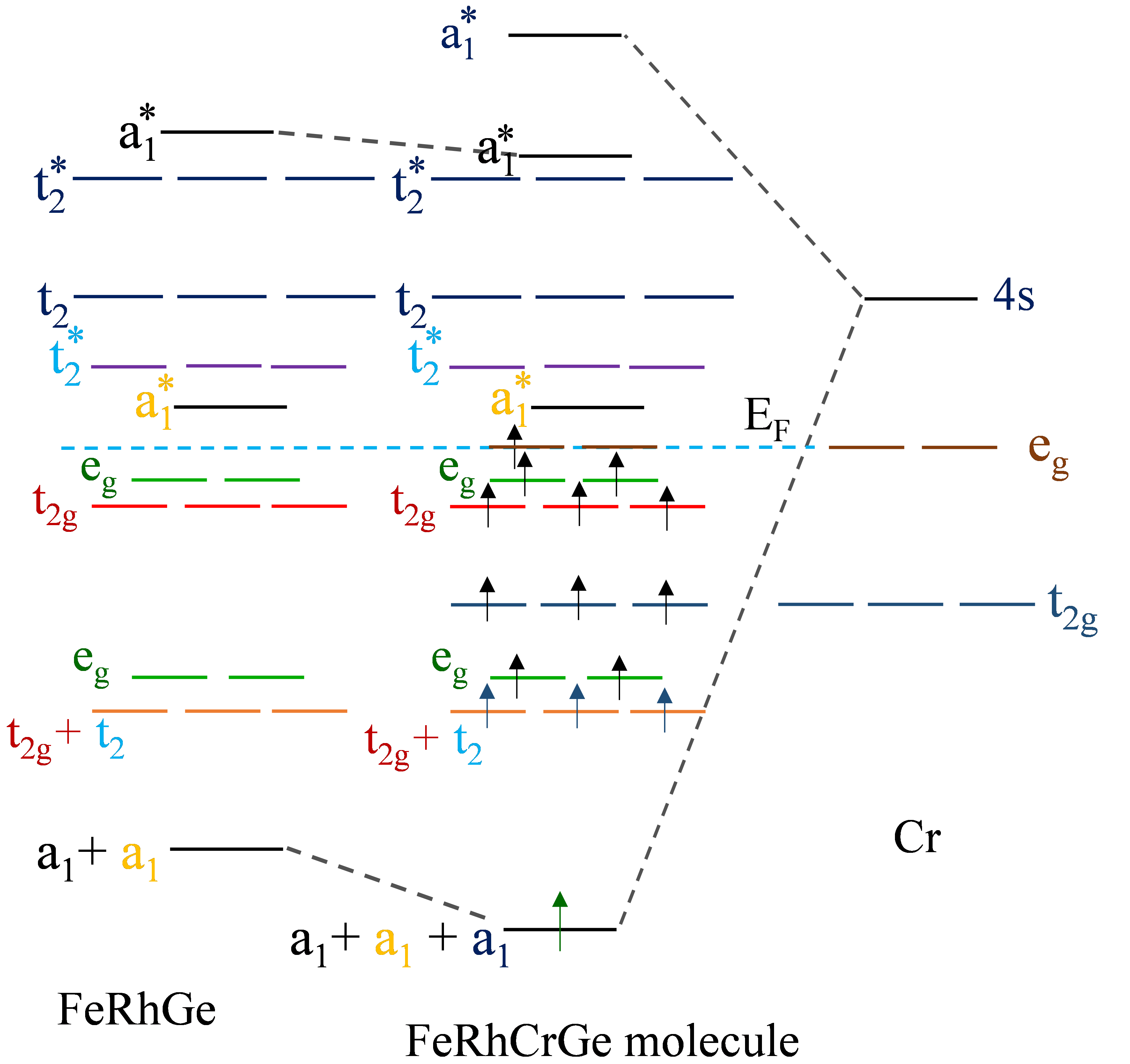}
\caption{Schematic view of spin up molecular energy levels for FeRhCrGe.}
\label{fig:schem-mol-up-FeRhCrGe}
\end{figure}

In the following, we explain the formation of unique band structure of FeRhCrGe using the molecular energy level diagram. Energy diagram at the molecular level is often quite informative in predicting the nature of bonding and transport. Such diagrams can be extracted from the orbital projected states at the $\Gamma$-point. Orbital decomposed atom-projected states for the spin up and down channels for FeRhCrGe are shown in Fig. S6 and S7 of SM.\cite{supp} The corresponding molecular level diagrams for various sub-molecules are shown in Fig. S8, S9 and S10.\cite{supp} Figure \ref{fig:schem-mol-up-FeRhCrGe} shows a schematic view of this diagram for spin up electrons. The energy levels are aligned as $a_1$, $t_{2g}'$, $e_g$, $t_{2g}$, $t_{2g}$, $e_g$, $e_g \ldots$ etc. for both spin bands. Here a$_1$, a$_{1}^{*}$, t$_2$ and t$_{2}^{*}$ are the bonding and anti-bonding orbitals having s and p characters respectively. $t_{2g}'$ represents the mixed character of triply degenerate $t_2$ \& $t_{2g}$ orbitals. $12$ energy levels are filled in spin down channel and E$_{\text{F}}$ lies in between anti-bonding levels of $t_{2g}$ $e_g$ of FeRh submolecule, whereas it lies at the degenerate $e_g$ level of octahedral site atoms in spin up channel (see Fig. \ref{fig:schem-mol-up-FeRhCrGe}). These $e_g$ orbitals at E$_F$ are well localized in energy and are partially filled. These partially filled orbitals are responsible for creating electron and hole pockets in bulk FeRhCrGe, thereby bringing unique spin semi-metallic property.

{\it \bf Conclusion :}  In summary, we have discovered a new class of spintronic materials, termed as spin semimetals. Using detailed experimental and theoretical investigations, we show that FeRhCrGe is a spin semimetal with unique properties. Furthermore, careful band structure calculations show that this alloy acquire additional features of type II Weyl semimetal and nodal line semi-metal located close to Fermi level for the spin up band. In conclusion, we confirm the co-existence of spin semimetallic and Weyl semimetallic behavior in FeRhCrGe. This makes it the first candidate material in this new class which is interesting both from the fundamental understanding as well as applied perspective.


{\it \bf Acknowledgments :}YV and SSS acknowledge the financial support provided by IIT Bombay. YV acknowledges the help of Mr. Akhilesh Kumar Patel and Miss Deepika Rani in the analysis of data. AA acknowledges National Center for Photovoltaic Research and Education (NCPRE), IIT Bombay for possible funding to support this research.


\bibliographystyle{apsrev4-1}
\bibliography{bib}

\end{document}